\documentclass[12pt]{article}
\usepackage{graphicx}

      \usepackage[cp1251]{inputenc} 

\begin{document}
 \noindent {\footnotesize\it Astronomy Letters, 2013, Vol. 39, No. 12, pp. 809--818.}
\newcommand{\dif}{\textrm{d}}

 \noindent
 \begin{tabular}{llllllllllllllllllllllllllllllllllllllllllllll}
 & & & & & & & & & & & & & & & & & & & & & & & & & & & & & & & & & & & & & \\\hline\hline
 \end{tabular}

 \vskip 1.0cm

 \centerline{\bf Galactic Rotation Curve and Spiral Density Wave Parameters from 73 Masers}
 \bigskip
 \centerline{\bf V.V. Bobylev$^{1,2},$ and A.T. Bajkova$^1$}
 \bigskip
 \centerline{\small $^{1}$\it Pulkovo Astronomical Observatory, St. Petersburg,  Russia}
 \centerline{\small $^{2}$\it Sobolev Astronomical Institute, St. Petersburg State University, Russia}
 \bigskip
 \bigskip

{\bf Abstract}—Based on kinematic data on masers with known
trigonometric parallaxes and measurements of the velocities of HI
clouds at tangential points in the inner Galaxy, we have refined
the parameters of the Allen-Santillan model Galactic potential and
constructed the Galactic rotation curve in a wide range of
Galactocentric distances, from 0 to 20~kpc. The circular rotation
velocity of the Sun for the adopted Galactocentric distance
$R_0=8$~kpc is $V_0=239\pm16$~km s$^{-1}$. We have obtained the
series of residual tangential, $\Delta V_{\theta}$, and radial,
$\Delta V_R,$ velocities for 73 masers. Based on these series, we
have determined the parameters of the Galactic spiral density wave
satisfying the linear Lin.Shu model using the method of
periodogram analysis that we proposed previously. The tangential
and radial perturbation amplitudes are $f_\theta=7.0\pm1.2$~km
s$^{-1}$ and $f_R=7.8\pm0.7$~km s$^{-1}$, respectively, the
perturbation wavelength is $\lambda=2.3\pm0.4$~kpc, and the pitch
angle of the spiral pattern in a two-armed model is
$i=-5.2^\circ\pm0.7^\circ$. The phase of the Sun $\chi_\odot$ in
the spiral density wave is $-50^\circ\pm15^\circ$ and
$-160^\circ\pm15^\circ$ from the residual tangential and radial
velocities, respectively.


\section*{INTRODUCTION}
A spectral (periodogram) analysis of the residual space velocities
for various young Galactic objects (hydrogen clouds, OB stars,
open star clusters, masers) tracing the spiral arms was used, for
example, by Clemens (1985), Bobylev et al. (2008), and Bobylev and
Bajkova (2010). In accordance with the model by Lin and Shu
(1964), such spiral density wave parameters as the pitch angle,
the perturbation amplitude and wavelength, and the phase of the
Sun in the spiral density wave were determined.

Bajkova and Bobylev (2012) studied the space velocities of 44
masers from the range of Galactocentric distances 3--14~kpc. An
amplitude of the velocity perturbations produced by the density
wave differing significantly from zero was found only in the
Galactocentric radial velocities of these sources,
$f_R=7.7\pm1.6$~km s$^{-1}$. In contrast, the amplitude of the
velocity perturbations in the residual rotation velocities
$f_\theta$ turned out to be close to zero, which looks strange.
For example, the residual rotation velocities of hydrogen clouds
and Cepheids show an amplitude of periodic perturbations
$f_\theta\approx6\pm1.5$~km s$^{-1}$ (Clemens 1985) and
$f_\theta=3.3\pm0.5$~km s$^{-1}$ (Bobylev and Bajkova 2012),
respectively. Theoretical estimates also suggest that $f_\theta$
should not differ greatly from $f_R$ (Burton 1971).

At present, there are data on about 80 Galactic masers that are
distributed in the range of Galactocentric distances 0–20 kpc; the
number of measured parallaxes constantly increases. Determining
consistent spiral density wave parameters obtained from both
radial and tangential velocities is of great interest. This
requires constructing a smooth rotation curve in a wide range of
distances maximally close to the data. A wide variety of methods
for determining a smooth rotation curve are known: constructing a
composite curve from the rotation velocities using polynomials of
various orders (Clemens 1985), solving Bottlinger’s kinematic
equations using several terms of the Taylor expansion of the
angular velocity of Galactic rotation (Zabolotskikh et al. 2002;
Popova and Loktin 2005; Bobylev et al. 2008), approximating the
rotation velocities by a set of exponential functions (Am\^ores et
al. 2009).

Here, we apply the dynamical method. It consists in constructing
the Galactic potential function. This approach is implemented, for
example, in the three-component model Galactic potential (Allen
and Santillan 1991; Khoperskov et al. 2013; Irrgang et al. 2013).

The goal of this paper is to refine the parameters of the
three-component model Galactic potential described by Allen and
Santillan (1991), to construct a smooth Galactic rotation curve in
the range 0--20 kpc, and, on this basis, to determine the spiral
density wave parameters from the residual tangential and radial
velocities of the maximum number of masers with accurately
measured trigonometric parallaxes known to date. To refine the
parameters of the Galactic potential and to construct the rotation
curve in a wide range of Galactocentric distances, 0--20~kpc, we
use both data on masers with Galactocentric distances from the
range 3--20~kpc and data on neutral hydrogen clouds at tangential
points in the inner Galaxy ($R<2$~kpc), where there are no
reliable maser velocity measurements at present.

\section*{DATA}
We use the maser coordinates, proper motions, and trigonometric
parallaxes measured by VLBI with an error, on average, less than
10\%. These masers are associated with very young objects,
protostars of mostly high masses (but there are also low-mass
ones; a number of massive supergiants are also known) located in
regions of active star formation, and, thus, trace well the spiral
arms.

One of the projects to obtain such data is the Japanese VERA (VLBI
Exploration of Radio Astrometry) project aimed at the observations
of Galactic H2O masers at 22~GHz (Hirota et al. 2007) and SiO
masers (there are very few such sources among young objects) at
43~GHz (Kim et al. 2008). Methanol (CH3OH) masers are observed at
12~GHz in the USA on VLBA (Reid et al. 2009a). The observations of
masers are also being carried out within the framework of the
European VLBI network, which now includes three Russian stations
of the QUASAR system (Rygl et al. 2010). The VLBI observations of
radio stars in continuum at 8.4~GHz are being carried out with the
same goals (Dzib et al. 2011).

Complete information about 44 masers (coordinates, line-of-sight
velocities, proper motions, and parallaxes) is presented in
Bajkova and Bobylev (2012). Subsequently, a number of papers with
new measurements have been published. New data on 31 sources are
presented in Table~1. Note that nine of them have no maser
emission, but they are fairly bright radio stars for which the
parallaxes were measured by VLBI with a high accuracy in
continuum. These include several low-mass stars in Taurus
(Hubble~4, V773 Tau AB, T Tau N, HDE 283572, and HP Tau/G2),
Serpens (EC 95), and Ophiuchus (S1 and DoAr21 Oph) as well as the
high-mass Xray binary Cyg X-1 with one of its components being a
black hole.

In the inner Galaxy $(R<2)$~kpc, as yet there are virtually now
data on masers. The trigonometric parallaxes (Reid et al. 2009b)
were measured only for two maser spots in the Sgr B2 region.
However, we do not use them for our studies and provide them only
for the purposes of illustration, because the spiral pattern of
the Galaxy begins from the bar ends ($R>3$~kpc). Therefore, our
final sample of sources based on which the spiral density wave
parameters are determined contains only 73 sources.

As a rule, invoking data on objects of various classes is required
to refine the parameters of the model Galactic potential and to
construct the rotation curve in a wide range of distances. For
example, apart from data on 30 masers, Irrgang et al. (2013) used
both data on hydrogen in the inner Galaxy ($R<2$~kpc) from Burton
and Gordon (1978) and data on CO in the middle Galactic region in
the range of distances 3--8~kpc from Clemens (1985) to construct
the Galactic potential. This was done because of the large ``gap''
in the distribution of masers. In our case, the situation is
distinctly different: we have already 73 masers that fill the
range of distances 3--14~kpc fairly densely; therefore, we use
only the line-of-sight velocities of neutral hydrogen clouds
located at tangential points in the inner Galaxy (Burton and
Gordon 1978).

      \begin{table}[]
      \begin{center}
      \caption{Initial data on the sources}
   \label{Tab0}
     {\small\protect
      \begin{tabular}{|l|r|r|r|r|r|r|r|r|r|r|r|}      \hline
      Source & $\alpha$ & $\delta$ & $\pi(\sigma_\pi)$ &
      $\mu^*_\alpha (\sigma_{\mu_\alpha})$ & $\mu_\delta(\sigma_{\mu_\delta})$ &
      $V_r(\sigma_{V_r})$ & Ref \\\hline

 IRAS 5168+36   & $ 80.0920$ & $ 36.6324$ & $ .532( .053)$ &$   .23(1.07)$ & $ -3.14( .28)$ & $-15.5( 1.9)$ &  (1) \\\hline
 NML Cyg        & $311.6064$ & $ 40.1165$ & $ .620( .047)$ &$ -1.55( .42)$ & $ -4.59( .41)$ & $  -.1( 2)  $ &  (2) \\\hline
 IRAS20143+36   & $304.0467$ & $ 36.7167$ & $ .368( .037)$ &$ -2.99( .16)$ & $ -4.37( .43)$ & $  7.0( 3)  $ &  (3) \\\hline
 PZ Cas         & $356.0137$ & $ 61.7895$ & $ .390( .022)$ &$ -3.20( .10)$ & $ -2.50( .10)$ & $-36.1( .7) $ &  (4) \\\hline
 IRAS22480+60   & $342.4953$ & $ 60.2991$ & $ .400( .025)$ &$ -2.58( .33)$ & $ -1.91( .17)$ & $-50.8( 3.5)$ &  (5) \\\hline
 RCW 122        & $260.0242$ & $-38.9603$ & $ .296( .026)$ &$  -.73( .04)$ & $ -2.83( .50)$ & $-12.6( 5)  $ &  (6) \\\hline
 Hubble 4       & $ 64.6960$ & $ 28.3354$ & $7.530( .030)$ &$  4.30( .05)$ & $-28.90( .30)$ & $  6.1( 1.7)$ &  (7) \\\hline
 HDE 283572     & $ 65.4952$ & $ 28.3018$ & $7.780( .040)$ &$  8.88( .06)$ & $-26.60( .10)$ & $  6.0( 1.5)$ &  (7) \\\hline
 T Tau N        & $ 65.4976$ & $ 19.5351$ & $6.820( .030)$ &$ 12.35( .04)$ & $-12.80( .05)$ & $  7.7( 1.2)$ &  (8) \\\hline
 V773 Tau AB    & $ 63.5538$ & $ 28.2034$ & $7.700( .190)$ &$  8.30( .50)$ & $-23.60( .50)$ & $  7.5(  .5)$ &  (9) \\\hline
 HP Tau/G2      & $ 68.9757$ & $ 22.9037$ & $6.200( .030)$ &$ 13.85( .03)$ & $-15.40( .20)$ & $  6.8( 1.8)$ & (10) \\\hline
 S1     Oph     & $246.6424$ & $-24.3912$ & $8.550( .500)$ &$ -3.88( .69)$ & $-31.55( .50)$ & $  3.0( 3)$ & (11) \\\hline
 DoAr21 Oph     & $246.5126$ & $-24.3934$ & $8.200( .370)$ &$-26.47( .92)$ & $-28.23( .73)$ & $  3.0( 3)$ & (11) \\\hline
 EC 95          & $277.4912$ & $  1.2128$ & $2.410( .020)$ &$   .70( .02)$ & $ -3.64( .10)$ & $  9.0( 3)$ & (12) \\\hline
 G074.03$-$1.71 & $306.2796$ & $ 34.8327$ & $ .629( .017)$ &$ -3.79( .18)$ & $ -4.88( .25)$ & $ 13.4( 3)$ & (13) \\\hline
 G075.76+0.33   & $305.4212$ & $ 37.4248$ & $ .285( .022)$ &$ -3.08( .06)$ & $ -4.56( .08)$ & $ -9.6( 3)$ & (13) \\\hline
 G075.78+0.34   & $305.4334$ & $ 37.4437$ & $ .281( .034)$ &$ -2.79( .07)$ & $ -4.72( .07)$ & $  3.4( 3)$ & (13) \\\hline
 G076.38$-$0.61 & $306.8562$ & $ 37.3801$ & $ .770( .053)$ &$ -3.73(3.00)$ & $ -3.84(3.0)$  & $  6.9( 3)$ & (13) \\\hline
 G079.87+1.17   & $307.6214$ & $ 41.2649$ & $ .620( .027)$ &$ -3.23(1.31)$ & $ -5.19(1.31)$ & $ -4.6( 3)$ & (13) \\\hline
 G090.21+2.32   & $315.5946$ & $ 50.0523$ & $1.483( .038)$ &$  -.67(3.13)$ & $  -.90(3.13)$ & $ -6.2( 3)$ & (13) \\\hline
 G092.67+3.07   & $317.3405$ & $ 52.3770$ & $ .613( .020)$ &$  -.69( .26)$ & $ -2.25( .33)$ & $ -3.7( 3)$ & (13) \\\hline
 G105.41+9.87   & $325.7769$ & $ 66.1153$ & $1.129( .063)$ &$  -.21(2.38)$ & $ -5.49(2.38)$ & $-12.1( 3)$ & (13) \\\hline
 IRAS20126+41   & $303.6084$ & $ 41.2257$ & $ .610( .020)$ &$ -4.14( .13)$ & $ -4.14( .13)$ & $ -4.0( 5)$ & (14) \\\hline
 W33 A f1       & $273.6435$ & $-17.8644$ & $ .408( .025)$ &$   .19( .08)$ & $ -2.52( .32)$ & $ 34.9( 5)$ & (15) \\\hline
 W33 A f2       & $273.6649$ & $-17.8668$ & $ .396( .032)$ &$  -.36( .08)$ & $ -2.22( .13)$ & $ 37.0( 5)$ & (15) \\\hline
 W33 Main f2    & $273.5576$ & $-17.9225$ & $ .343( .037)$ &$  -.60( .11)$ & $  -.99( .13)$ & $ 34.1( 5)$ & (15) \\\hline
 G012.88+0.48   & $272.9646$ & $-17.5247$ & $ .340( .036)$ &$   .12( .13)$ & $ -2.66( .23)$ & $ 29.4( 5)$ & (15) \\\hline
 IRAS05137+39   & $ 79.3073$ & $ 39.3722$ & $ .086( .027)$ &$   .30( .10)$ & $  -.89( .27)$ & $-26.0( 3)$ & (16) \\\hline
 Cyg X-1        & $299.5903$ & $ 35.2016$ & $ .539( .033)$ &$ -3.78( .06)$ & $ -6.40( .12)$ & $ 15.5( 5)$ & (17) \\\hline
 SgrB2N         & $266.8330$ & $-28.3720$ & $ .128( .015)$ &$  -.32( .05)$ & $ -4.69( .11)$ & $ 64.0( 5)$ & (18) \\\hline
 SgrB2M         & $266.8340$ & $-28.3845$ & $ .130( .012)$ &$ -1.23( .04)$ & $ -3.84( .11)$ & $ 61.0( 5)$ & (18) \\\hline
      \end{tabular}}
      \end{center}
{\small
 Note. $\alpha$ and $\delta$ in deg., $\pi$ in mas,
 $\mu^*_\alpha=\mu_\alpha\cos\delta$ and $\mu_\delta$ in mas/yr,
 $V_r=V_r(LSR)$ in km/s,
 The numbers mark the references to papers: (1) Sakai et al. (2012); (2) Zhang et al. (2012); (3) Yamaguchi et al. (2012);
(4) Kusuno et al. (2012); (5) Imai et al. (2012); (6) Wu et al.
(2012); (7) Torres et al. (2007); (8) Loinard et al. (2007); (9)
Lestrade et al. (1999); (10) Torres et al. (2009); (11) Loinard et
al. (2008); (12) Dzib et al. (2010); (13) Xu et al. (2013); (14)
Moscadelli et al. (2011); (15) Immer et al. (2013); (16) Honma et
al. (2011); (17) Reid et al. (2011); (18) Reid et al. (2009b).
        }
      \end{table}

\section*{METHODS AND APPROACHES}
The values of two quantities should be known to determine the
circular velocities of stars: the Galactocentric distance of the
Sun $R_0,$ which we take to be 8~kpc, and the circular rotation
velocity of the solar neighborhood $V_0.$

\subsection*{Determining the Velocity $V_0$}
To determine the velocity $V_0=R_0|\Omega_0|$ from observational
data, we use the equations derived from Bottlinger’s formulas with
the angular velocity of Galactic rotation $\Omega$ expanded in a
series to terms of the second order of smallness in $r/R_0:$
 \begin{equation}
 \begin{array}{lll}
 V_r=-u_\odot\cos b\cos l-v_\odot\cos b\sin l-w_\odot\sin b+\\
 +R_0(R-R_0)\sin l\cos b \Omega'_0+0.5R_0 (R-R_0)^2 \sin l\cos b \Omega''_0,
 \label{EQ-1}
 \end{array}
 \end{equation}
 \begin{equation}
 \begin{array}{lll}
 V_l= u_\odot\sin l-v_\odot\cos l+(R-R_0)(R_0\cos l-r\cos b)\Omega'_0+\\
  +(R-R_0)^2 (R_0\cos l - r\cos b)0.5\Omega''_0 - r \Omega_0 \cos b,
 \label{EQ-2}
 \end{array}
 \end{equation}
 \begin{equation}
 \begin{array}{lll}
 V_b=u_\odot\cos l \sin b+v_\odot\sin l \sin b-w_\odot\cos b-\\
    -R_0(R-R_0)\sin l\sin b\Omega'_0-0.5R_0(R-R_0)^2\sin l\sin b\Omega''_0,
 \label{EQ-3}
 \end{array}
 \end{equation}
where $V_r$ is the line-of-sight velocity of the star (in ~km
s$^{-1}$); $r=1/\pi$ is the heliocentric distance of the star;
$V_l=4.74 r \mu_l\cos b$ and $V_b=4.74 r \mu_b$ are the proper
motion velocity components of the star (in mas yr$^{-1}$) in the
$l$ and $b$ directions, respectively; the coefficient 4.74 is the
quotient of the number of kilometers in an astronomical unit by
the number of seconds in a tropical year;
$u_\odot,v_\odot,w_\odot$ are the stellar group velocity
components relative to the Sun taken with the opposite sign (the
velocity $U$ is directed toward the Galactic center, $V$ is in the
direction of Galactic rotation, $W$ is directed to the north
Galactic pole); $R_0$ is the Galactocentric distance of the Sun;
$\Omega_0$ is the angular velocity of rotation at the distance
$R_0;$ the parameters $\Omega'_0$ and $\Omega''_0$ are the first
and second derivatives of the angular velocity, respectively; the
distance of the star to the Galactic rotation axis R is calculated
from the formula
 \begin{equation}
 \begin{array}{lll}
 R^2=r^2\cos^2 b-2R_0 r\cos b\cos l+R^2_0.
 \label{RR}
 \end{array}
 \end{equation}
Equations (1) are solved by the least-squares method with weights
of the form
 \begin{equation}
 \begin{array}{lll}
 w_r=S_0/\sqrt {S_0^2+\sigma^2_{V_r}},\quad
 w_l=\beta^2 S_0/\sqrt {S_0^2+\sigma^2_{V_l}},\quad
 w_b=\gamma^2 S_0/\sqrt {S_0^2+\sigma^2_{V_b}},
 \label{WESA}
 \end{array}
 \end{equation}
where $S_0$ is the ``cosmic'' dispersion taken to be 8~km
s$^{-1}$; $\sigma_{V_r},$ $\sigma_{V_l},$ and $\sigma_{V_b}$ are
the errors in the corresponding observed velocities;
$\beta=\sigma_{V_r}/\sigma_{V_l}$ and
$\gamma=\sigma_{V_r}/\sigma_{V_b}$ are the scale factors that we
determined using data on open star clusters (Bobylev et al. 2007),
$\beta=1$ and $\gamma=2$.

\subsection*{Determining the Radial and Tangential Velocities}
For hydrogen clouds, there is only the line-of-sight velocity
$V_r$. The projection of the circular rotation velocity $V_\theta$
is calculated from the well-known formula (Burton 1971)
 \begin{equation}
  V_{\theta}=|R\Omega_0|+RV_r/(R_0\sin l\cos b).
 \label{rot-1}
 \end{equation}
Note that radio astronomers usually give the line-of-sight
velocities relative to the local standard of rest (this is true
for both masers and the line-of-sight velocities of HI clouds).
Therefore, they should be reduced to the heliocentric frame of
reference (in Table~1, the line-of-sight velocities of all stars
are given relative to the local standard of rest, LSR). This
requires using the parameters of the standard solar motion
 \begin{equation}
  \begin{array}{rll}
 (\alpha,\delta)_{1900}=(270^\circ, 30^\circ),~V=20~\hbox{km s$^{-1}$}, \\
 (U,V,W)_{LSR}=(10.3,15.3,7.7)~\hbox{km s$^{-1}$}. \label{standart}
  \end{array}
 \end{equation}
Since the hydrogen clouds are assumed to be located at tangential
points (in the first or fourth quadrants), the following simple
relations hold for them:
 \begin{equation}
  \begin{array}{rll}
     R&=& R_0|\sin l|, \\
     r&=& R_0 \cos l.
   \label{ff-1}
  \end{array}
 \end{equation}
The spatial coordinates of all objects are calculated in a
rectangular $X, Y, Z$ coordinate system. The components of the
observed space velocities $U$ and $V$ are calculated from the
projections $V_r,$ $V_l,$ and $V_b;$ these are used to find two
projections: $V_R$ directed radially away from the Galactic center
and $V_\theta$ orthogonal to it:
 \begin{equation}
  \begin{array}{rll}
     V_{R}&=&-U\cos \theta+(V_0+V)\sin \theta, \\
  V_{\theta}&=& U\sin \theta+(V_0+V)\cos \theta,
   \label{ff-2}
  \end{array}
 \end{equation}
where the object’s position angle $\theta$ is defined as
$\tan\theta=Y/(R_0-X).$ The velocities $U$ and $V$ are assumed to
be free from the solar velocity relative to the centroid
$V_\odot(u_\odot,v_\odot,w_\odot).$

\subsection*{The Model Galactic Potential}
Here, we use the model Galactic potential by Allen and Santillan
(1991). The axisymmetric Galactic potential is represented as the
sum of three components--the central (bulge), disk, and halo ones:
 \begin{equation}
 \Phi = \Phi_C + \Phi_D + \Phi_H.
 \label{pot-1}
 \end{equation}
The central component of the Galactic potential in cylindrical
coordinates ($r,\theta,z$) is represented in the form proposed by
Miyamoto and Nagai (1975):
 \begin{equation}
 \Phi_C=-\frac{M_C}{(r^2+z^2+b_C^2)^{1/2}},
 \label{pot-2}
 \end{equation}
where $M_C$ is the mass, $b_C$ is the scale length. The disk
component is represented as the Miyamoto--Nagai (1975) disk:
\begin{equation}
\Phi_D=-\frac{M_D}{\{r^2+[a_D+(z^2+b_D^2)^{1/2}]^2\}^{1/2}},
 \label{pot-3}
\end{equation}
where $M_D$ is the mass, $a_D,$ and $b_D$ are the scale lengths.
The halo component is represented in the form proposed by Allen
and Martos (1986):
 \begin{equation}
 \Phi_H=-\frac{M(R)}{R}-\int_R^{100}{{\frac{1}{R^{'}}}{\frac{dM(R^{'})}{dR^{'}}}}dR^{'},
 \label{pot-4}
 \end{equation}
where
$$
M(R)=\frac{M_H(R/a_H)^{2.02}}{1+(R/a_H)^{1.02}},
$$
where $M_H$ is the mass, ah is the scale length. If $R$ is
measured in kpc and $M_C,M_D,M_H$ are measured in units of the
Galactic mass (M$_G$), $2.32\times10^7 M_\odot$, then the
gravitational constant $G=1$ and the unit of measurement for the
potential $\Phi$ and its individual components (2)--(4) is
100~km$^2$ s$^{-2}.$

\subsection*{The Rotation Curve}
The tangential velocities are expressed in terms of the Galactic
potential components as
 \begin{equation}
 V^2_{\theta} = R \left\{ {{\partial\Phi_C}\over {\partial R}}+
  {{\partial\Phi_D}\over {\partial R}}+ {{\partial\Phi_H}\over {\partial R}}\right\},
 \label{RotCurve-45}
 \end{equation}
Substituting Eqs.~(2)--(4) into (5) and setting $z=0,$ we will
obtain an analytical expression for the smooth rotation curve:
 \begin{equation}
 V_{\theta} =  \left\{ {{R^2 M_C}\over {(R^2+b^2_C)^{3/2}}} +
               {{R^2 M_D}\over {(R^2+(a_D+b_D)^2 )^{3/2}}}+
               {{R^{1.02} M_H}\over {a_H^{2.02}(1+(R/a_H)^{1.02} )} }
               \right\}^{1/2}.
\label{RotCurve-55}
 \end{equation}
The parameters of the model potential are determined by
least-squares fitting to the measured velocities of the masers and
hydrogen clouds. The residual tangential velocities $\Delta
V_{\theta}$ used below to determine the spiral density wave
parameters are found as the differences of the tangential
velocities $V_{\theta}$ and the smooth rotation curve~(6).

\subsection*{Estimating the Spiral Density Wave Parameters}
To take into account the influence of the spiral density wave, we
used the simplest kinematic model based on the linear density wave
theory by Lin and Shu (1964), in which the potential perturbation
is in the form of a traveling wave. Then,
 \begin{equation}
 \begin{array}{lll}
   V_R= -f_R \cos \chi, \\
  \Delta V_{\theta}= f_\theta \sin \chi,
 \label{VR-Vtheta}
 \end{array}
 \end{equation}
Here, $V_R$ and $\Delta V_{\theta}$ are the radial and tangential
velocity perturbations produced by the spiral density wave, $f_R$
and $f_\theta$ are the amplitudes of the radial and tangential
velocity perturbations, $i$ is the spiral pitch angle ($i<0$ for
winding spirals), $m$ is the number of arms (here, we take $m=2$),
the wave phase $\chi$ is
 \begin{equation}
   \chi=m[\cot (i)\ln (R/R_0)-\theta]+\chi_\odot,
 \label{chi-creze}
 \end{equation}
where $\chi_\odot$ is the radial phase of the Sun in the spiral
density wave; we measure this angle (following Rohlfs 1977) from
the center of the Carina--Sagittarius spiral arm
($R\approx7$~kpc).

The parameter $\lambda$, the distance (along the Galactocentric
radial direction) between adjacent spiral arm segments in the
solar neighborhood (the wavelength of the spiral density wave), is
calculated from the relation
\begin{equation}
 \frac{2\pi R_0}{\lambda} = m\cot(i).
 \label{a-04}
\end{equation}
Let there be a series of measured velocities $V_n(R_n)$ (these can
be both radial, $V_R,$ and residual tangential, $\Delta
V_{\theta}$, velocities) at points with Galactocentric distances
$R_n$ and position angles $\theta_n$, $n=1,\dots,N$ where $N$ is
the number of objects. The objective of our spectral analysis is
to extract a periodicity from the data series in accordance with
the model (7)--(9) describing a spiral density wave with
parameters $f_R (f_\theta), \lambda (i),$ and $\chi_\odot$.

Having taken into account the logarithmic character of the spiral
density wave and the position angles of the objects $\theta_n$,
our spectral (periodogram) analysis of the series of velocity
perturbations is reduced to calculating the square of the
amplitude (power spectrum) of the standard Fourier transform
(Bajkova and Bobylev 2012):
\begin{equation}
 \bar{V}_{\lambda_k} = \frac{1} {N}\sum_{n=1}^{N} V^{'}_n(R^{'}_n)
 \exp\Bigl(-j\frac {2\pi R^{'}_n}{\lambda_k}\Bigr),
 \label{29}
\end{equation}
where $\bar{V}_{\lambda_k}$ is the $k$th harmonic of the Fourier
transform, with the wavelength being $\lambda_k=D/k$, $D$ is the
period of the series being analyzed,
\begin{equation}
 R^{'}_{n}=R_{\circ}\ln(R_n/R_{\circ}),
 \label{21}
\end{equation}
$$
 V^{'}_n(R^{'}_n)=V_n(R^{'}_n)\times\exp(jm\theta_n).
$$
The algorithm of searching for periodicities modified to properly
determine not only the wavelength but also the amplitude of the
perturbations is described in detail in Bajkova and Bobylev
(2012).

Obviously, the sought-for wavelength $\lambda$ corresponds to the
peak value of the power spectrum $S_{peak}$. The pitch angle of
the spiral density wave can be derived from Eq. (9). We determine
the perturbation amplitude and phase by fitting the harmonic with
the wavelength found to the observational data. The following
relation can also be used to estimate the perturbation amplitude:
$$
f_R(f_\theta)=\sqrt{4\times S_{peak}}.
$$

\begin{figure}[t]
{\begin{center}
 \includegraphics[width=0.9\textwidth]{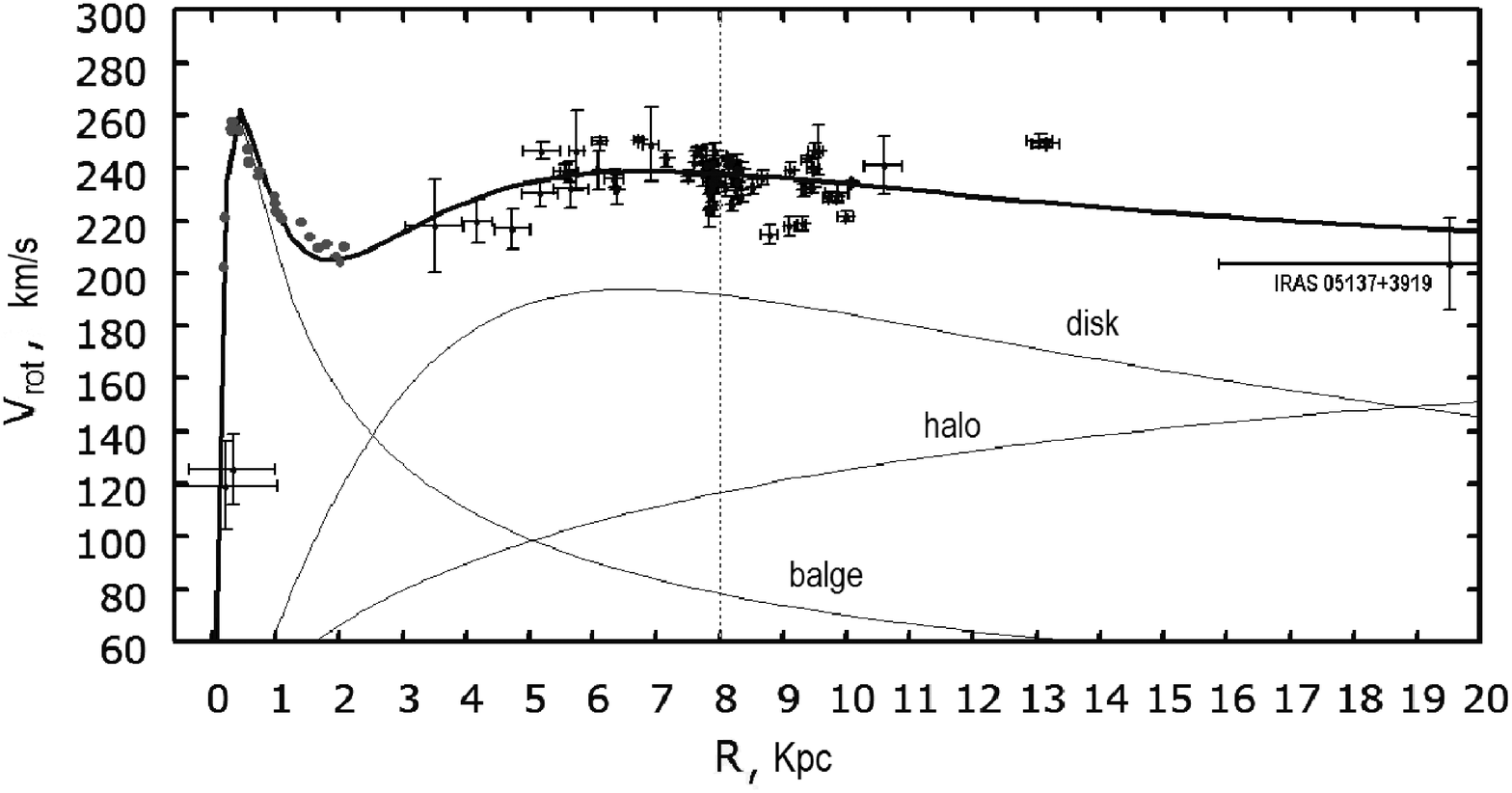}
 \caption{
Galactic rotation curve (thick line), the thin lines indicate the
contribution from the bulge, disk, and halo to the rotation curve,
the dotted line marks the position of the Sun, the gray circles
indicate the data on neutron hydrogen (Burton and Gordon 1978),
the dots with error bars indicate the data on masers.}
 \label{f1}
\end{center}}
\end{figure}

\section*{RESULTS}
\subsection*{The Velocity $V_0$}
We obtained several solutions of the system of equations (1) by
the least-squares method with various constraints on the sample
radius $(r)$ and the parallax error $(e_\pi/\pi)$. These solutions
differ insignificantly, but it should be noted that $\Omega_0$
increases with decreasing sample radius from $\Omega_0=28$~km
s$^{-1}$ kpc$^{-1}$ for all masers to $\Omega_0=32$~km s$^{-1}$
kpc$^{-1}$ for the nearest ones.

For example, based on the entire sample of masers, we obtained the
following solution:
 \begin{equation}
 \begin{array}{lll}
 (u_\odot,v_\odot,w_\odot)= (7.4,16.3,8.3)\pm(1.2,1.1,1.0)~\hbox {km s$^{-1}$},        \\
 \Omega_0      = 28.7 \pm0.6~\hbox{km s$^{-1}$ kpc$^{-1}$},  \\
 \Omega^{'}_0  = -4.11\pm0.13~\hbox{km s$^{-1}$ kpc$^{-2}$}, \\
 \Omega^{''}_0 =  0.676\pm0.057~\hbox{km s$^{-1}$ kpc$^{-3}$},
 \label{solution-01}
 \end{array}
 \end{equation}
with the error per unit weight being $\sigma_0=8.11$~km s$^{-1}$.

Based on a sample of 55 masers from the range $r<3.5$~kpc with
$e_\pi/\pi<10\%,$ we obtained the following solution:
 \begin{equation}
 \begin{array}{lll}
 (u_\odot,v_\odot,w_\odot)=  (8.8,17.3,7.9)\pm(1.4,1.3,1.0)~\hbox {km s$^{-1}$},
 \label{solution-1UVW}
 \end{array} \end{equation} \begin{equation} \begin{array}{lll}
 \Omega_0      = 29.9 \pm1.1~\hbox{km s$^{-1}$ kpc$^{-1}$},        \\
 \Omega^{'}_0  = -4.27\pm0.20~\hbox{km s$^{-1}$ kpc$^{-2}$}, \\
 \Omega^{''}_0 =  0.915\pm0.166~\hbox{km s$^{-1}$ kpc$^{-3}$}.
 \label{solution-1}
 \end{array}
 \end{equation}
In this case, the error per unit weight is
$\sigma_0=7.65$~km~s$^{-1}$, which is considerably smaller than
that in the previous case. Therefore, we adopted the solution for
$\Omega_0$ obtained from 55 masers with more accurate parallaxes.
Since $\Omega_0$ is a local parameter, the solution obtained from
nearer objects with more accurate parallaxes may be considered to
be more correct than that obtained from all objects, including the
distant ones. The sought-for circular velocity of the Sun will
then be $V_0=R_0|\Omega_0|=239\pm16$~km s$^{-1}$.

The rotation velocities ($V_{\theta}$) of the hydrogen clouds and
masers calculated with the velocity $V_0$ found are presented in
Fig.~1. We took into account the group motion of the stars
relative to the Sun (11). It can be seen that the velocities (11),
especially $v_\odot$, differ from the LSR parameters
$(U_\odot,V_\odot,W_\odot)_{LSR}=(11.1,12.2,7.3)$~km s$^{-1}$
found by Sch\"onrich et al. (2010). In our opinion (Bobylev and
Bajkova 2010), such a difference is due to the influence of the
spiral density wave. Based on a sample of 30 masers from the Orion
arm, Xu et al. (2013) found a difference $\Delta
v_\odot\approx5$~km s$^{-1}$ compared to the velocity from
Sch\"onrich et al. (2010).

Note that the Galactic rotation curve constructed with the
parameters (12) is applicable only in a limited solar neighborhood
with a radius of 4--6 kpc and goes upward at large $R>14$~kpc
(Bobylev and Bajkova 2010). In such an approach, for example, the
residual velocity of the maser IRAS~05137$+$3919 will be very
high. Thus, the rotation curve is currently difficult to construct
by expanding the angular velocity (1) into a series because of the
very small amount of data at large distances $R.$ Therefore, here
we use an approach based on a refinement of the Galactic potential
parameters to construct the rotation curve.

\begin{table}[t]
 \caption{Parameters of the model Galactic potential}
\begin{center}
  \label{Tab1}
\begin{tabular}{|c|r|c|}\hline
 $M_C$ &   493 M$_G$   \\\hline
 $M_D$ &  4599 M$_G$   \\\hline
 $M_H$ &  6526 M$_G$   \\\hline
 $b_C$ & 0.2815 kpc    \\\hline
 $a_D$ & 4.4555 kpc    \\\hline
 $b_D$ & 0.25 kpc      \\\hline
 $a_H$ & 15.9 kpc      \\\hline
\end{tabular}
\end{center}
\end{table}

\subsection*{The Galactic Potential and Rotation Curve}
The refined parameters of the model Galactic potential found by
fitting Eq.~(5) (by the least-squares method) to the measured
velocities of the masers and hydrogen clouds are given in Table~2.
The Galactic rotation curve and the contributions from each of the
three potential components to the total velocity obtained from
Eq.~(6) are shown in Fig.~1. The masses $M_C, M_D,$ and $M_H$ from
Table~2 were multiplied by 100.

\subsection*{The Spiral Density Wave Parameters}
In Fig. 2, the radial velocities $V_R$ of the masers and their
residual rotation velocities $\Delta V_{\theta}$ are plotted
against the distance $R'$ (Eq. (10)), and Fig. 3 shows their power
spectra. The significance level of the main peak is 0.999 for the
radial velocities (Fig. 3a) and 0.86 for the rotation velocities
(Fig. 3b). As a result, we obtained the following spiral density
wave parameters based on our sample of 73 masers independently for
each of the velocity components ($V_R$ and $\Delta V_{\theta}$):
 \begin{equation}
 \begin{array}{lll}
                 f_R= 7.8\pm0.7~\hbox{km s$^{-1}$}, \\
            f_\theta= 7.0\pm1.2~\hbox{km s$^{-1}$}, \\
         \lambda_{R}= 2.4\pm0.4~\hbox{kpc},  \\
    \lambda_{\theta}= 2.3\pm0.4~\hbox{kpc},  \\
      (\chi_\odot)_R=-160^\circ\pm15^\circ,  \\
  (\chi_\odot)_\theta=-50^\circ\pm15^\circ.
 \label{solution-2}
 \end{array}
 \end{equation}
The pitch angle for the model of a two-armed pattern is
$i=-5.2^\circ\pm0.7^\circ$.

\begin{figure}[t]
{\begin{center}
 \includegraphics[width=0.84\textwidth]{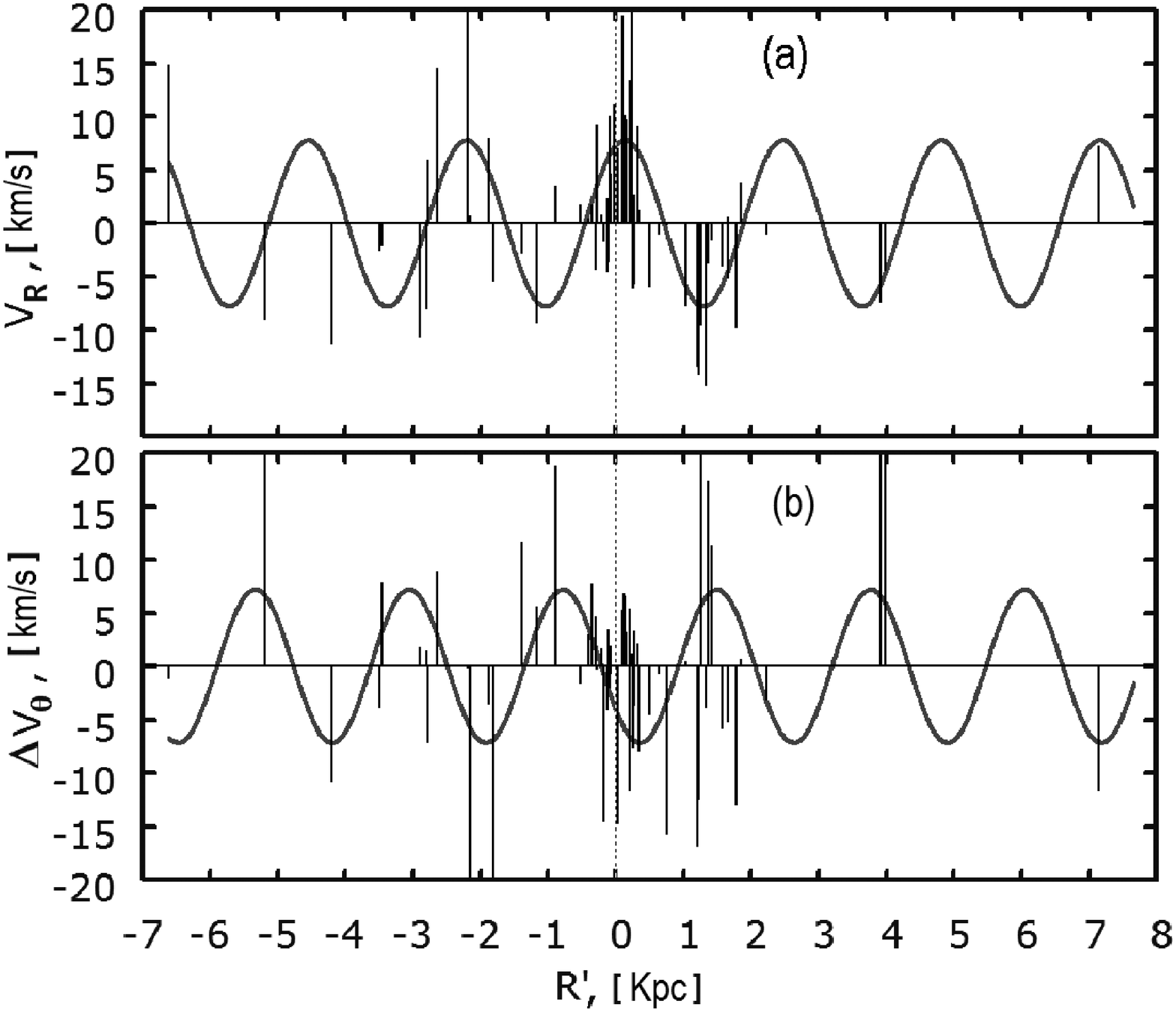}
 \caption{
(a) Galactocentric radial velocities $V_R$ of the masers versus
distance $R'$; (b) their residual rotation velocities $\Delta
V_{\theta}$. The dotted line marks the position of the Sun. }
 \label{Rad-Thet-f2}
\end{center}}
\end{figure}
\begin{figure}[t]
{\begin{center}
 \includegraphics[width=0.99\textwidth]{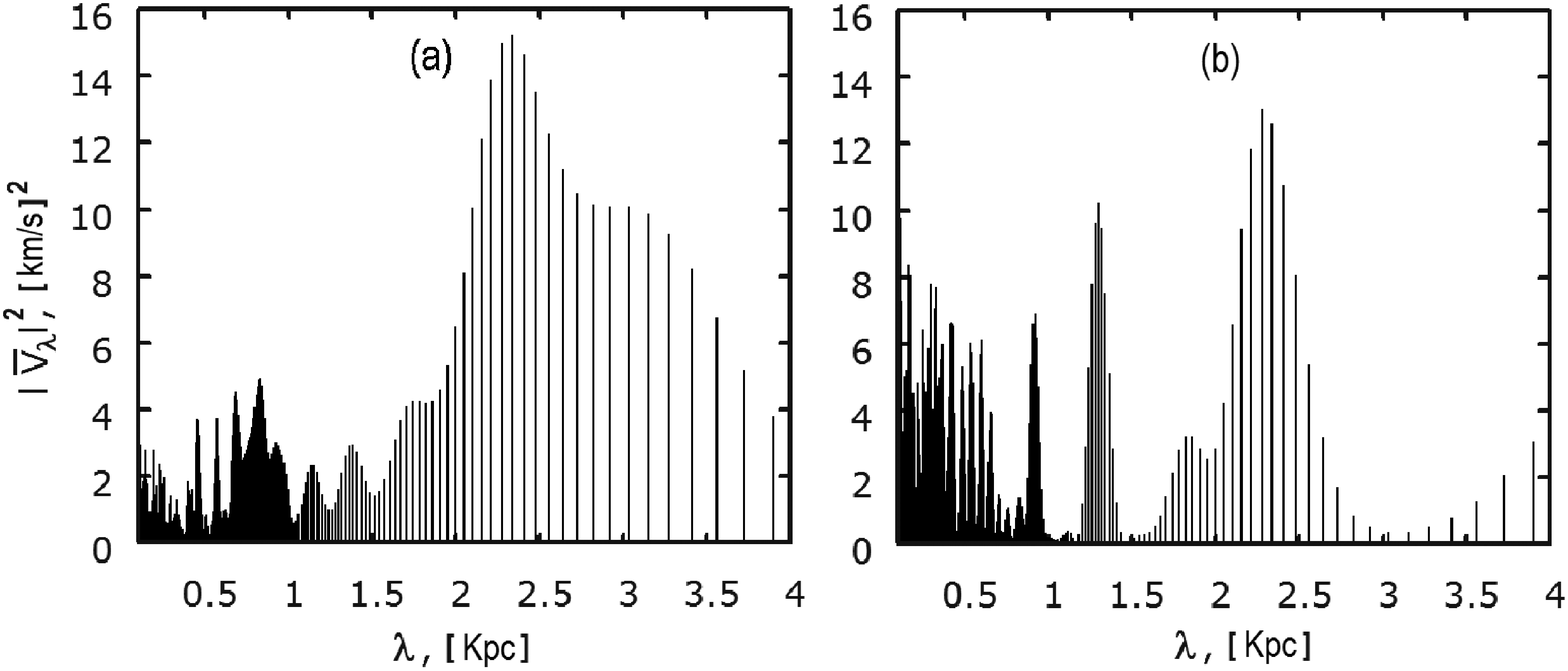}
 \caption{
(a) Power spectrum for the radial velocities $V_R$ of the masers;
(b) power spectrum for their residual rotation velocities $\Delta
V_{\theta}$. }
 \label{Spectr-f3}
\end{center}}
\end{figure}

\section*{DISCUSSION}
\subsection*{Parameters of the Galactic Rotation Curve}
Based on 52 Galactic masers, Honma et al. (2012) found the angular
velocity of Galactic rotation $\Omega_0=31.09\pm0.78$~km s$^{-1}$
kpc$^{-1}$ and the linear velocity $V_0=R_0|\Omega_0|=238\pm14$~km
s$^{-1}$ (at $R_0=8.05\pm0.45$~kpc found). The parameters (12) are
in good agreement with these results. Therefore, it should be
noted that the masers belong to the most rapidly rotating fraction
of the Galactic disk. This is not surprising, because they are the
youngest objects that have recently been formed from hydrogen.

Other authors found a high angular velocity of their Galactic
rotation even from a smaller number of masers. Based on a sample
of 18 masers, Reid et al. (2009a) found $\Omega_0=30.3\pm0.9$~km
s$^{-1}$ kpc$^{-1}$. Based on the same sample, McMillan and Binney
(2010) showed that $\Omega_0$ lies within the range 29.9--31.6~km
s$^{-1}$ kpc$^{-1}$ at various $R_0$ and obtained an estimate of
$V_0 = 247\pm19$~km s$^{-1}$ for $R_0=7.8\pm0.4$~kpc. Based on a
sample of 18 masers, Bovy et al. (2009) found $V_0=244\pm13$~km
s$^{-1}$ for $R_0=8.2$~kpc. Based on a sample of 18 masers,
Brunthaler et al. (2011) found $V_0=246\pm7$~km s$^{-1}$ and
$R_0=8.3\pm0.23$~kpc. Using 28 masers, Stepanishchev and Bobylev
(2011) obtained an estimate of $V_0=243\pm16$~km s$^{-1}$
($R_0=8.0\pm0.5$~kpc). The parameters of the Galactic rotation
curve (12) found are in good agreement with the results of
analyzing other young objects of the Galactic disk rapidly
rotating around the center. For example, based on OB associations,
Mel'nik and Dambis (2009) found $\Omega_0 =31\pm1$~km s$^{-1}$
kpc$^{-1}$; based on a sample of blue supergiants, Zabolotskikh et
al. (2002) obtained $\Omega_0=29.6\pm1.6$~km s$^{-1}$ kpc$^{-1}$
and $\Omega'_0=-4.76\pm0.32$~km s$^{-1}$ kpc$^{-2}$; based on OB3
stars, Bobylev and Bajkova (2001) found
 $\Omega_0 = 31.5\pm0.9$~km s$^{-1}$ kpc$^{-1}$,
 $\Omega^{'}_0 = -4.49\pm0.12$~km s$^{-1}$ kpc$^{-2}$ and
 $\Omega^{''}_0 = 1.05\pm0.38$~km s$^{-1}$ kpc$^{-3}.$
Sch\"onrich (2012) obtained an estimate of $V_0=250\pm9$ km
s$^{-1}$ for $R_0=8.27\pm0.29$~kpc from data on $\approx$220 000
stars from the SEGUE (Sloan Extension for Galactic Understanding
and Exploration) catalogue.

\subsection*{The Galactic Potential}
Irrgang et al. (2013) considered three model Galactic potentials
constructed from data on hydrogen clouds and masers. In all cases,
the velocity V0 was found to be close to 240 km s$^{-1}$ (for
$R_0\approx8.3$~kpc). The potentials differ by the expressions
describing the halo contribution. In particular, model~I, just as
in our case, is a refinement of the model by Allen and Santillan
(1991). Note that Irrgang et al. (2013) used both HI data at
tangential points from Burton and Gordon (1978) and CO data from
Clemens (1985). The rotation curve for model I from Irrgang et al.
(2013) in the region 4.5~kpc passes closer to the CO observations
above the masers, although, on the whole, the results turned out
to be close. However, with regard to the approach of the rotation
curve to the data on masers in the entire range, our curve
(Fig.~1) fits better the data. This turned out to be important for
analyzing the periodic perturbations produced by the influence of
the spiral density wave.

\subsection*{Spiral Density Wave Parameters}

Based on the radial $(V_R)$ velocities of 44 masers, Bajkova and
Bobylev (2012) obtained the following parameters related to the
influence of the spiral density wave: the perturbation amplitude
$f_R=7.7^{+1.7}_{-1.5}$~km s$^{-1}$, the perturbation wavelength
$\lambda=2.2^{+0.4}_{-0.1}$~kpc, the pitch angle of the spiral
density wave for the model of a two-armed pattern $(m=2)$
$i=-5^{+0.2^\circ}_{-0.9^\circ},$ and the Sun's phase in the
spiral density wave $\chi_\odot=-147^{+3^\circ}_{-17^\circ}$. The
parameters~(13) are in good agreement with these results.

Here, for the first time we have found parameters of the
perturbations produced by the spiral density wave in the residual
rotation velocities ($\Delta V_{\theta}$) of masers that differ
significantly from zero. Previously, it was possible to reveal
such periodicities in the residual rotation velocities of hydrogen
clouds (Clemens 1985) or middle-age Cepheids (Bobylev and Bajkova
2012). As can be seen from solution (13) and Fig.~2, the
perturbation amplitude and wavelength found from both radial and
tangential velocities are in good agreement between themselves.
There is a significant discrepancy only in the Sun's phases,
$\Delta\chi_\odot=110^\circ,$ in the spiral density wave. The
value of $\chi_\odot$ obtained from the radial velocities is more
trustworthy, because it agrees better with the position of the Sun
in the spiral pattern that we found previously (Bobylev and
Bajkova 2013) by an independent method from our analysis of the
spatial distribution of masers, $\chi_\odot=-152^\circ$.

\section*{CONCLUSIONS}

Based on kinematic data on 73 masers with known trigonometric
parallaxes measured by VLBI and measurements of the velocities of
HI clouds at tangential points in the inner Galaxy ($R<2$~kpc), we
refined the parameters of the Allen.Santillan (1991) model
Galactic potential and, on this basis, constructed the Galactic
rotation curve in a wide range of Galactocentric distances, from 0
to 20 kpc. The circular rotation velocity of the Sun for the
adopted Galactocentric distance $R_0=8$~kpc was found to be
$V_0=239\pm16$~km s$^{-1}$.

Based on the series of residual tangential, $\Delta V_{\theta}$,
and radial, $V_R,$ velocities for masers, we determined the
parameters of the Galactic spiral density wave satisfying the
linear Lin--Shu model using the method of periodogram analysis
that we proposed previously. The tangential and radial
perturbation amplitudes are $f_\theta=7.0\pm1.2$~km s$^{-1}$ and
$f_R=7.8\pm0.7$~km s$^{-1}$, respectively, the perturbation
wavelength is $\lambda=2.3\pm0.4$~kpc, and the pitch angle of the
spiral pattern in a two-armed model is $i=-5.2^\circ\pm0.7^\circ$.
The phase of the Sun $\chi_\odot$ in the spiral density wave is
$-50^\circ\pm15^\circ$ and $-160^\circ\pm15^\circ$ from the
residual tangential and radial velocities, respectively.

\subsection*{ACKNOWLEDGMENTS} We are grateful to the referee for
helpful remarks that contributed to a improvement of the paper.
This work was supported by the ``Nonstationary Phenomena in
Objects of the Universe'' Program of the Presidium of the Russian
Academy of Sciences and the ``Multiwavelength Astrophysical
Research'' grant no. NSh--16245.2012.2 from the President of the
Russian Federation. In our work, we widely used the SIMBAD
astronomical database.

\section*{REFERENCES}
{\small
 \quad~~1. C. Allen and M. A. Martos, Rev. Mex. Astron. Astrofis. 13, 137 (1986).

2. C. Allen and A. Santillan, Rev. Mex. Astron. Astrofis. 22, 255
(1991).

3. E.B. Am\^ores, J.R.D. L\'epine, and Yu.N. Mishurov, Mon. Not.
R. Astron. Soc. 400, 1768 (2009).

4. A.T. Bajkova and V.V. Bobylev, Astron. Lett. 38, 549 (2012).

5. V.V. Bobylev and A.T. Bajkova, Mon. Not. R. Astron. Soc. 408,
1788 (2010).

6. V.V. Bobylev and A.T. Bajkova, Astron. Lett. 37, 526 (2011).

7. V.V. Bobylev and A.T. Bajkova, Astron. Lett. 38, 638 (2012).

8. V.V. Bobylev and A.T. Bajkova, Astron. Lett. 39, 532 (2013).

9. V.V. Bobylev, A.T. Bajkova, and A. S. Stepanishchev, Astron.
Lett. 34, 515 (2008).

10. J. Bovy, D.W. Hogg, and H.-W. Rix, Astrophys. J. 704, 1704
(2009).

11. A. Brunthaler, M.J. Reid, K.M. Menten, et al., Astron. Nachr.
332, 461 (2011).

12. W.B. Burton, Astron. Astrophys. 10, 76 (1971).

13. W.B. Burton and M. A. Gordon, Astron. Astrophys. 63, 7 (1978).

14. D.P. Clemens, Astrophys. J. 295, 422 (1985).

15. S. Dzib, L. Loinard, A.J. Mioduszewski, et al., Astrophys. J.
718, 610 (2010).

16. S. Dzib, L. Loinard, L.F. Rodriguez, et al., Astrophys. J.
733, 71 (2011).

17. T. Hirota, T. Bushimata, Y.K. Choi, et al., Publ. Astron. Soc.
Jpn. 59, 897 (2007).

18. M. Honma, T. Hirota, Y. Kan-ya, et al., Publ. Astron. Soc.
Jpn. 63, 17 (2011).

19. M. Honma, T. Nagayama, K. Ando, et al., Publ. Astron. Soc.
Jpn. 64, 136 (2012).

20. H. Imai, N. Sakai, H. Nakanishi, et al., Publ. Astron. Soc.
Jpn. 64, 142 (2012).

21. K. Immer, M.J. Reid, K.M. Menten, et al., Astron. Astrophys.
553, 1171 (2013). 

22. A. Irrgang, B. Wilcox, E. Tucker, et al., Astron. Astrophys.
549, 137 (2013).

23. S.A. Khoperskov, E.O. Vasiliev, A.M. Sobolev, et al., Mon.
Not. R. Astron. Soc. 428, 2311 (2013).

24. M.K. Kim, T. Hirota, M. Honma, et al., Publ. Astron. Soc. Jpn.
60, 991 (2008).

25. K. Kusuno, A. Yoshiharu, I. Hiroshi, et al., in Proceedings of
the East Asia VLBI Workshop 2012, May 30--Jun. 2, 2012 (2012).

26. J.-F. Lestrade, R.A. Preston, D.L. Jones, et al., Astron.
Astrophys. 344, 1014 (1999).

27. C.C. Lin and F.H. Shu, Astrophys. J. 140, 646 (1964).

28. L. Loinard, R.M. Torres, A.J. Mioduszewski, et al., Astrophys.
J. 671, 546 (2007).

29. L. Loinard, R.M. Torres, A.J. Mioduszewski, et al., Astrophys.
J. 671, L29 (2008).

30. P.J. McMillan and J.J. Binney, Mon. Not. R. Astron. Soc. 402,
934 (2010).

31. A.M. Mel’nik and A.K. Dambis, Mon. Not. R. Astron. Soc. 400,
518 (2009).

32. M. Miyamoto and R. Nagai, Publ. Astron. Soc. Jpn. 27, 533
(1975).

33. L. Moscadelli, R. Cesaroni, M.J. Rioja, et al., Astron.
Astrophys. 526, A66 (2011).

34. M.E. Popova and A.V. Loktin, Astron. Lett. 31, 663 (2005).

35. M.J. Reid, K.M. Menten, X.W. Zheng, et al., Astrophys. J. 700,
137 (2009a).

36. M.J. Reid, K.M. Menten, X.W. Zheng, et al., Astrophys. J. 705,
1548 (2009b).

37. M.J. Reid, J.E. McClintock, R. Narayan, et al., Astrophys. J.
742, 83 (2011).

38. K. Rohlfs, Lectures on Density Wave Theory (Springer, Berlin,
1977).

39. K.L.J. Rygl, A. Brunthaler, M.J. Reid, et al., Astron.
Astrophys. 511, A2 (2010).

40. M. Sakai, M. Honma, H. Nakanishi, et al., Publ. Astron. Soc.
Jpn. 64, 108 (2012).

41. R. Sch\"onrich, Mon. Not. R. Astron. Soc. 427, 274 (2012).

42. R. Sch\"onrich, J. Binney, and W. Dehnen, Mon. Not. R. Astron.
Soc. 403, 1829 (2010).

43. A.S. Stepanishchev and V.V. Bobylev, Astron. Lett. 37, 254
(2011).

44. R.M. Torres, L. Loinard, A.J. Mioduszewski, et al., Astrophys.
J. 671, 1813 (2007).

45. R.M. Torres, L. Loinard, A.J. Mioduszewski, et al., Astrophys.
J. 698, 242 (2009).

46. Y.W. Wu, Y. Xu, K.M. Menten, et al., in Cosmic Masers—from
$OH$ to $H_0,$ Proceedings of the IAU Symposium No. 287, Ed. by
R.S. Booth, E.M.L. Humphreys, and W.H. T. Vlemmings (Cambridge,
2012), p. 425.

47. Y. Xu, J.J. Li, M.J. Reid, et al., Astrophys.~J. 769, 15 (2013). 

48. Y. Yamaguchi, T. Handa, T. Omodaka, et al., in Proceedings of
the East Asia VLBI Workshop 2012, May 30–Jun. 2, 2012 (2012).

49. M.V. Zabolotskikh, A.S. Rastorguev, and A.K. Dambis, Astron.
Lett. 28, 454 (2002).

50. B. Zhang, M.J. Reid, K.M. Menten, et al., Astron. Astrophys.
544, A42 (2012).

}

\end{document}